\documentclass{nime-alternate}
\usepackage[utf8]{inputenc} 
\usepackage[T1]{fontenc}    
\usepackage[hidelinks]{hyperref}       
\usepackage{url}
\usepackage{booktabs}       
\usepackage{amsfonts}       
\usepackage{nicefrac}       
\usepackage{microtype}      
\usepackage{graphicx}

\usepackage{anonymize}

\conferenceinfo{NIME'19,}{June 3-6, 2019, Federal University of Rio Grande do Sul, ~~~~~~  Porto Alegre,  Brazil.}

\title{An Interactive Musical Prediction System with Mixture Density
  Recurrent Neural Networks}

\label{key}

\numberofauthors{2}

\author{
  \alignauthor
  \anonymize{Charles P. Martin} \\
  \affaddr{\anonymize{Department of Informatics and RITMO}}\\
  \affaddr{\anonymize{University of Oslo, Norway}}\\
  \email{\anonymize{charlepm@ifi.uio.no}} \\
  \alignauthor
  \anonymize{Jim Torresen}\\
  \affaddr{\anonymize{Department of Informatics and RITMO}}\\
  \affaddr{\anonymize{University of Oslo, Norway}}\\
  \email{\anonymize{jimtoer@ifi.uio.no}}
}

\begin{document}

\maketitle



\begin{abstract}
This paper is about creating digital musical instruments where a predictive neural network model is integrated into the interactive system. Rather than predicting symbolic music (e.g., MIDI notes), we suggest that predicting future control data from the user and precise temporal information can lead to new and interesting interactive possibilities. We propose that a mixture density recurrent neural network (MDRNN) is an appropriate model for this task. The predictions can be used to fill-in control data when the user stops performing, or as a kind of filter on the user's own input. We present an interactive MDRNN prediction server that allows rapid prototyping of new NIMEs featuring predictive musical interaction by recording datasets, training MDRNN models, and experimenting with interaction modes. We illustrate our system with several example NIMEs applying this idea. Our evaluation shows that real-time predictive interaction is viable even on single-board computers and that small models are appropriate for small datasets.
\end{abstract}




\section{Introduction}

In this paper, we consider how mixture density recurrent neural
networks (MDRNNs)~\cite{Bishop:1994aa, Graves:2013aa} can be applied
to real-time gestural prediction in new interfaces for musical
expression (NIMEs) and present an interactive system for training and
applying MDRNNs to a broad range of NIMEs. Research applying deep
learning to music \emph{generation} is rapidly appearing, but few of
these systems have been applied in the service of real-time musical
\emph{performance}. We feel that deep neural networks (DNNs) can
extend creative possibilities for NIME performers and designers;
however, these users need DNN models that are appropriate for typical
NIME-data, and better tools to allow rapid-prototyping and creative
exploration with these models.

Present work in musical AI is usually focussed on high-level symbolic
music; however, most NIME control data is better represented as
low-level continuous sensor values. We propose to use MDRNNs to model
this data, including time-deltas between each reading. This approach
has the advantage of modelling music at the
embodied~\cite{Leman:2018aa} control level; such models imitate
performing on instruments, not composing music. Another advantage is
in representing rhythms absolutely---as a sequence of real-valued
time-deltas---rather than being limited to a sixteenth-note grid.
MDRNNs have previously been applied to control data in
sketching~\cite{Ha:2017ab} and handwriting~\cite{Graves:2013aa}, both
creative tasks.

\begin{figure}
  \centering
  \includegraphics[width=0.5\textwidth]{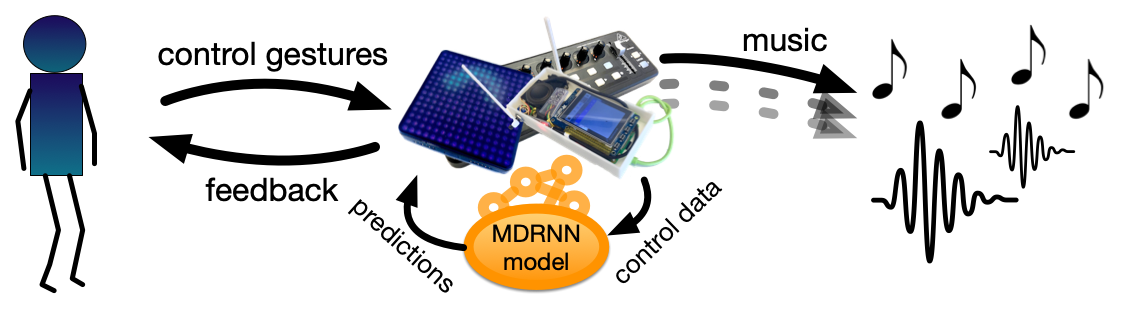}
  \caption{Our interactive musical prediction system allows NIMEs to
    be integrated with a mixture density recurrent neural network
    (MDRNN) that can support or accompany a human performance.}
  \label{fig:predictive-interaction}
\end{figure}

We have developed an interactive musical prediction system (IMPS), to
accelerate development of musical MDRNN models that predict musical
control data in real-time during performance (see Figure
\ref{fig:predictive-interaction}). This system assists with data
collection, model training, and real-time inference. IMPS connects to
typical NIME software and devices via a simple open sound control
(OSC) interface and can apply an MDRNN through a number of new
predictive interaction paradigms to accompany or support performers.
This tool provides a new solution for MDRNN-based NIME development for
artists and computer musicians. We imagine that artists could train
MDRNN models on small datasets of interaction data, tailored to
commercial or DIY interfaces applied in their practice. While small
models may not represent all possible musical interactions, they might
perform well enough to imitate aspects of an artist's style.

The main contributions of this research are the novel IMPS system that
assists with data-collection, training, and real-time application of
MDRNNs. We discuss the design of this system, and in particular the
advantages of an MDRNN model over other popular DNNs. Our evaluation
is focussed on the training and application of this system in real-time
performance. We describe a number of example NIMEs developed using these
tools, and show that incorporation of real-time MDRNN predictions is
feasible on even single-board computers such as the Raspberry Pi.

\begin{figure*}
    \centering
    \includegraphics[width=0.99\textwidth]{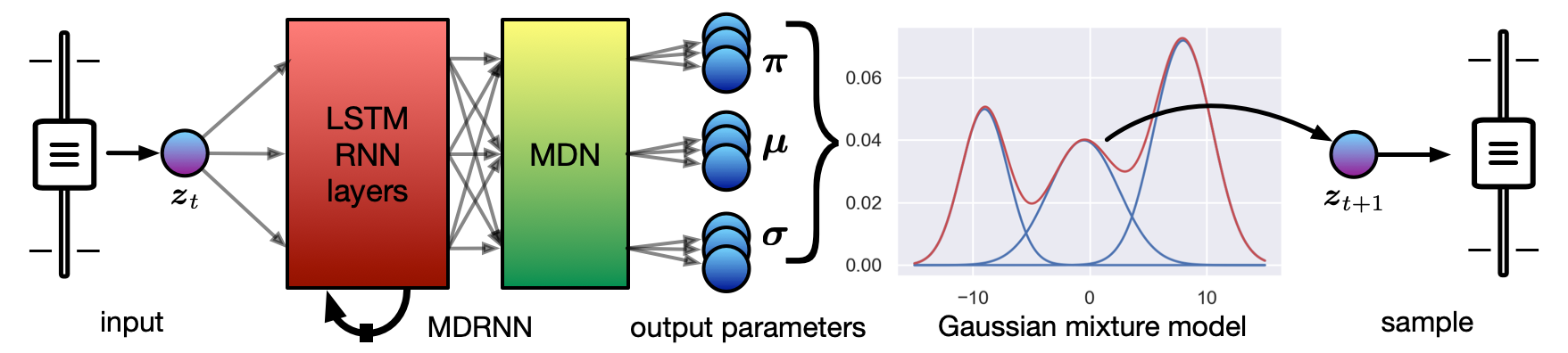}
    \caption{An MDRNN transforms the output of an RNN to form the
      parameters of a Gaussian mixture model. Our system allows a
      configurable MDRNN to be applied to predicting musical control data, such
      as mixer sensors or touchscreen data.}
    \label{fig:mdn-example}
\end{figure*}

\section{Machine learning in NIMEs}

Machine learning (ML) has been a part of NIME designs since at least
the early 1990s~\cite{Lee:1991aa}. More recently, the adoption of ML
algorithms into NIMEs has accelerated due to frameworks such as
Wekinator~\cite{fiebrink2009metainstrument}, SARC EyesWeb
Catalogue~\cite{Gillian:2011fu} or ml.lib~\cite{Bullock:2015qr}, that
allow training and application of ML algorithms through a GUI or
computer music tools such as Max or Pd.

ML has typically been applied in NIMEs for two tasks: mapping from
gesture to high-level synthesiser commands, and modelling of ongoing
musical processes~\cite{Martin:2018dpmim}. On the mapping side,
classification algorithms and shallow NNs have been useful for mapping
multidimensional control data such as wind instrument
keys~\cite{jsnyder1:2014} to musical notes or high level gestures.
Modelling algorithms such as Markov models~\cite{Pachet:2003wd} or
factor oracles~\cite{Levy:2012aa}, have been used to develop systems
that learn to imitate musical styles, or interact with an improvising
musician. These modelling systems usually operate on high-level
musical notes or acoustic phrases, and as a result have not been as
applicable to data from the multiple continuous controllers of many
NIMEs.

\subsection{Deep Neural Networks for Music}

In recent years much research on musical applications of DNNs has
appeared (see \cite{Sturm:2017aa} for a survey). Artificial neural
networks (ANNs) process data by applying simple mathematical
transformations grouped into units, inspired by neurons in the brain.
The strength of connections between units, how much data is allowed to
move between them, are the parameters of an ANN. By arranging units
deeply, into layers, ANNs be taught to accomplish high-level tasks
such as recognising images and generating musical melodies.

For music applications, multi-layered recurrent neural networks (RNNs)
with long short-term memory (LSTM) units are often used to learn
sequences of symbolic music data. In RNNs, data can be stored
in between computations and fed back as extra input. This means that
RNNs can learn to understand temporal relationships between sequences
of data that are applied to their inputs~\cite{Goodfellow:2016aa}.
LSTM units are RNN components that contain an additional memory state
and four internal gates that control how data is stored, released,
retained and ``forgotten'' from the memory state.

Typical LSTM music models generate music by predicting one discrete
note symbol at a time using MIDI-~\cite{Jaques:2017aa} or
text-inspired representations~\cite{Sturm:2017aa}. Rather than
directly predicting a note, the outputs of these networks form a
categorical (softmax) probability distribution from which one discrete option
can be sampled. The temperature (or diversity) of this distribution
can be adjusted to favour the highest scoring option, to boost the
chance of sampling less likely notes.

Symbolic LSTM RNNs can be used to model discrete musical data (e.g.,
notes from a piano keyboard on a 16th-note rhythmic grid), but a
different architecture is required to directly predict
\emph{continuous-valued} musical data such as input from touch
sensors. To model touchscreen data, Hantrakul~\cite{Hantrakul:2018aa}
used a RNN without a stochastically sampled output. Mixture density
networks (MDN) have been applied to touchscreen
performances~\cite{robojam2018}, which preserves stochastic sampling.
That system added direct prediction of rhythm enabling pauses of
arbitrary length, taps, as well as continual swirls. As will be further
explained below, MDNs are highly applicable to many types of
continuous control data, not just touchscreens.

While tools to define and train DNNs (e.g., Keras and TensorFlow) have
become more accessible, the lack of a NIME-focussed toolkit has held
back the application of DNNs within an interactive music context. This
research attempts to fill this gap with an MDRNN prediction system
that can be explored without additional programming.

\subsection{Mixture Density RNNs}

A mixture density network (MDN) transforms the outputs of a neural
network into the parameters of a Gaussian mixture model
(GMM)~\cite{Bishop:1994aa}, as shown in Figure \ref{fig:mdn-example}.
Such a model ``mixes'' a number of Gaussian (or normal) distributions
(in blue) with weights corresponding to the likelihood of each
components, to form a more complex distribution (in red). This allows
the model to represent phenomena that appear to be drawn from multiple
 distributions. As shown in Figure \ref{fig:mdn-example}, the
MDN's output parameters consist of centres ($\mu$) and scales
($\sigma$) for each component distribution, and a weight ($\pi$) for
each component. An MDN can be applied to the outputs of an RNN forming
an MDRNN that can be trained on sequential data such as 2D pen
movements for handwriting~\cite{Graves:2013aa} and
sketches~\cite{Ha:2017ab}, as well as musical touchscreen
data~\cite{robojam2018}.

In a creative process such as musical improvisation, multiple choices
for the next note to perform or action to take could be artistically
valid. This suggests that some kind of multi-modal
distribution, such as a GMM, would be appropriate to accurately model
such a process. An MDRNN thus forms a useful network for regression
problems involving multiple correct answers, or that require a certain
amount of stochasticity when sampling, as in creative tasks.

One of the complexities of an MDN is the error (loss) function, used
for training, which is derived from the probability density function of
the mixture model. While this is straightforward for a
1-dimensional~\cite{Bishop:1994aa} or 2D~\cite{Graves:2013aa} case,
mixtures of higher-dimensional Gaussian distributions are more
complex. Our system, outlined below, makes use of recent features of
TensorFlow to generate a tractable loss function for arbitrary many
dimensions of data.


\section{System Design}

Our predictive musical interaction system consists of two components:
a Keras implementation of an MDRNN with an MDN layer that extends to
sequential data of arbitrary dimension and a set of Python
applications to facilitate data collection, model training, and
real-time interactive prediction. The system is open
source\footnote{IMPS Code: \anonymize{https://doi.org/10.5281/zenodo.2580175}} and
operated through a command line interface that enables these tasks to
be performed within the context of a NIME prototyping and performance
process. This system is designed to allow artists, as well as machine
learning researchers, to apply MDRNNs to creative work and other
applications.

\subsection{MDN Layer}

Our MDN layer allows the selection of the number of mixture components
and the dimension of the data to be modelled. Each mixture component
is a multivariate Gaussian distribution limited to a diagonal
covariance matrix. So for $K$ components and dimension $N$, for one
prediction, the MDN generates $K$ mixing coefficients
($\boldsymbol{\pi}$), $K \times N$ means ($\boldsymbol{\mu}$) and
$K \times N$ scales ($\boldsymbol{\sigma}$, diagonals of the
covariance matrix). This is illustrated for $N=1$, $K=3$ in
Figure \ref{fig:mdn-example} with an example of the resulting mixture
distribution.

The limited covariance matrix means that the loss function can be
easily calculated using TensorFlow Probability's \texttt{Mixture},
\texttt{Categorical}, and \texttt{MultivariateNormalDiag} functions.
The loss function for an MDN relies on the number of mixture
components as well of the dimension of each component, so this
function is generated on demand for these parameters.

\subsection{MDRNN}

We provide an abstraction, \texttt{PredictiveMusicMDRNN}, for
generating an appropriate MDRNN for learning musical sequences. This
is constructed using multiple layers of LSTM units followed by an MDN
layer. The number of layers, LSTM units, mixture components, and
distribution dimensions are configurable hyperparameters, but we
suggest that 2-layer networks with 5 mixture components are used. The
abstraction assumes that the first dimension of the data will be used
for the time since the previous sample ($dt$) and will thus be
positive and nonzero. The other dimensions of the data are assumed to
be between 0 and 1. This means that to model data with 2 continuous
control variables ($x_1$ and $x_2$), a 3D MDRNN is required ($dt$,
$x_1$, $x_2$). Our abstraction can construct networks for training
(unrolled to the correct training sequence length) and for inference
(with sequence length set to 1, and LSTM state stored in between
calculations), and provides a function to make inferences one at a
time as input data becomes available.

We provide a high-level control of the size of the MDRNN which allows
the user to trade potential learning capacity for speed of training
and inference. Four presets are provided that change the number of LSTM
units as follows: s--64, m--128, l--256, xl--512. The `s' network has
around 50K parameters while the `xl' network has 3M. Further
customisation to hyperparameters is easily made through a command line interface.

\subsection{Predictive Musical Interaction Controller}


The main IMPS application receives interaction messages from a control
interface over OSC (e.g., \texttt{/interface}, $x_1$, $x_2$, \ldots).
All interaction messages are automatically logged to CSV files in
order to build up a training dataset.

IMPS' MDRNN module can accept interface messages as input, performing
inference and sampling to produce predictions of the user's next
control interaction (as illustrated in Figure \ref{fig:mdn-example}).
Information from the MDRNN's input conditions the LSTM layers'
internal states, which means that predictions are influenced by
previous interactions. The MDRNN can also perform predictions on top
of predictions by connecting output to input, allowing it to
generate new control signals based on previous experiences.
Predictions are stored internally and sent back to the interface (with
the address \texttt{/prediction}), at the time indicated by the $dt$
values output by the MDRNN. This allows the MDRNN to perform
arbitrarily timed rests and rhythms.






The input and output of the MDRNN can be connected in different ways
to messages from the interface and the outgoing stream of predicted
control values. Given that messages may arrive faster than MDRNN
predictions are made, IMPS uses concurrent queues to keep up as much
as possible. We have defined four interaction modes in IMPS to explore
these configurations in performance:
\begin{description}
\setlength{\itemsep}{0pt}
\setlength{\parskip}{0pt}
\item[No Predictions.] Interface messages condition the MDRNN memory,
  but the output predictions are ignored.

\item[Filter.] Interface messages condition the MDRNN memory; output
  predictions are played as quickly as possible.

\item[Call-and-Response.] The MDRNN is conditioned when the user
  plays; if the user stops, the MDRNN generates
  new control signals until they interact again.

\item[Battle.] The user and MDRNN play together, with the MDRNN
  continually generating control signals regardless of the user's
  actions.

\end{description}
These predictive interactions are a starting point for further
research. In particular, they demonstrate that our system is capable
of both mapping (through the filter interaction), and modelling
(through call-and-response).


\section{Example Applications}

In this section we present several NIMEs that integrate our MDRNN
predictive model to demonstrate its flexibility for a variety of
interactive music modalities. These systems allow the output from the
MDRNN to be sonified as well as represented visually or physically.
The systems have differing degrees of freedom, from one single
controller, to an eight-knob interface. These systems are illustrated
in our video abstract\footnote{Video Abstract:
  \anonymize{https://doi.org/10.5281/zenodo.2597494}}.




\subsection{EMPI}

\begin{figure}
    \centering
    \includegraphics[width=0.3\textwidth]{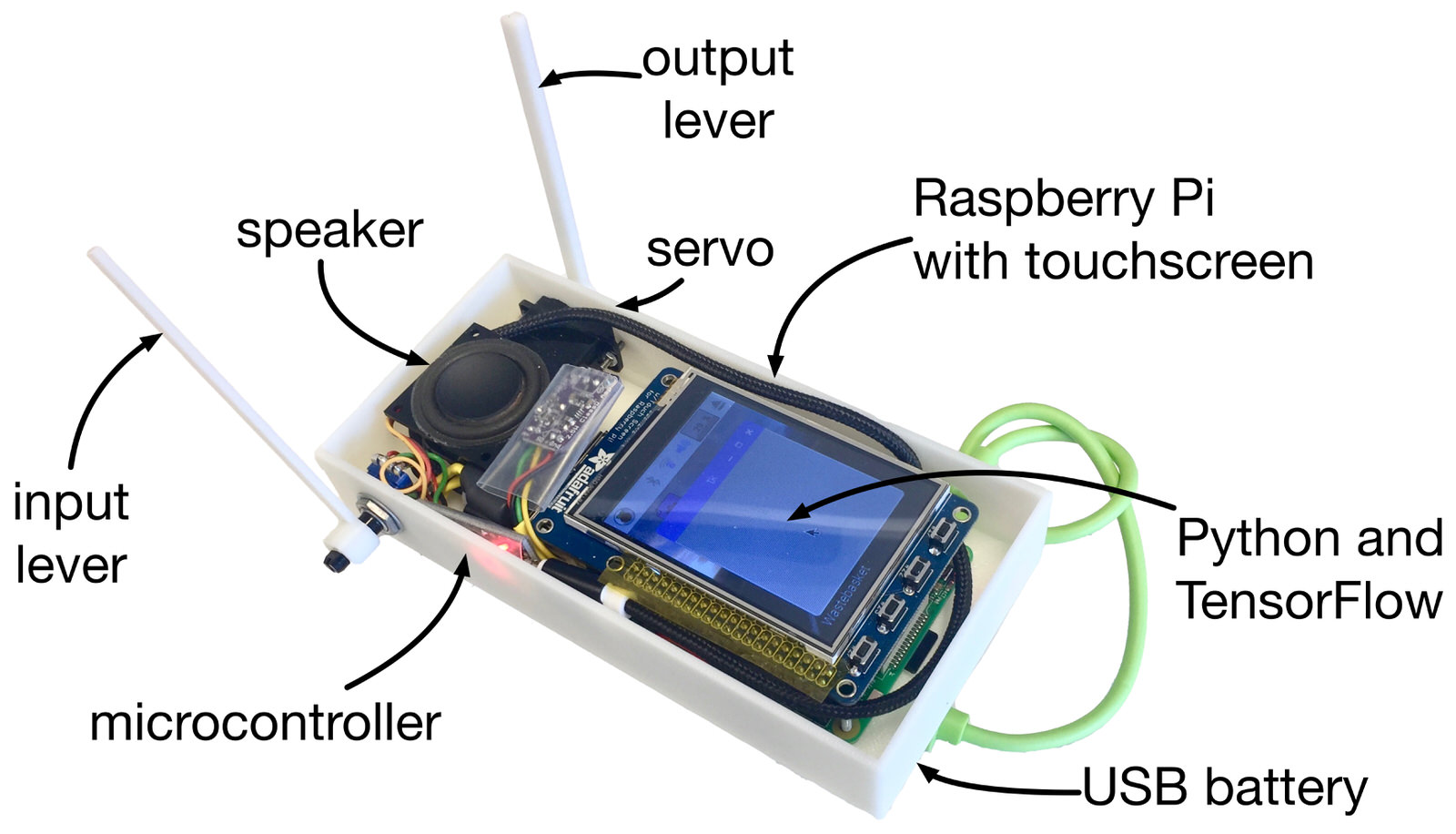}
    \caption{The EMPI self-contained NIME (enclosure open) represents
      predictions physically and sonically. It uses a 2D MDRNN ($dt$,
      input value).}
    \label{fig:empi}
\end{figure}

The Embodied Musical Predictive Interface (EMPI) is a self-contained
NIME with a single dimension of continuous input and output through
two physical levers, a Raspberry Pi, touchscreen, and speaker. EMPI
was designed to explore the simplest predictive musical interactions:
where one dimension of input is modelled along with time. The MDRNN
model was trained on a 10-minute human performance with the input
lever. The EMPI demonstrates that even a Raspberry Pi can be used for
predictions from a small MDRNN in a real-time situation. In
performance, the predictions are sonified through the EMPI's speaker
as well as physically represented through the servo-controlled lever,
this imbues the NIME with a sense of independent agency.

\subsection{LightPad Block}
\label{sec:lightpad-block}

\begin{figure}
  \centering
  \includegraphics[width=0.20\textwidth]{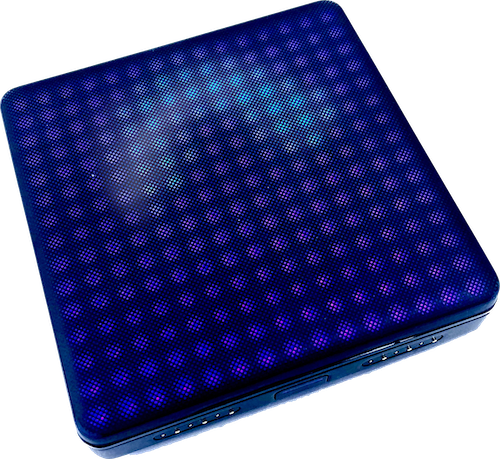}
  \caption{The LightPad is a pressure-sensitive multitouch controller
    with an LED display. It can be used with 4D MDRNN model ($dt$,
    $x$, $y$, pressure)}
  \label{fig:lightpad}
\end{figure}

The Roli LightPad Block (see Figure \ref{fig:lightpad}) is a small
multitouch controller with a flexible pressure sensitive surface and
an LED display. The LightPad's expressive inputs and the ability to
visually represent these inputs through the display make it ideal for
experimenting with predictive interaction. We developed a Pd patch
that communicates between our predictive interaction system (OSC) and
the LightPad (MIDI over Bluetooth). This allows MIDI messages from the
LightPad, and predictions from the MDRNN to be mapped to software
synthesisers. The LightPad's LED display is usually used to illuminate
the user's touches as well as the state of music software. We feed
MDRNN predictions back to the LightPad which are displayed in a
different colour. This allows an intuitive view of the user's actions
and predictive responses during performance.

The LightPad data was applied to a 4D MDRNN ($dt$, $x$-position,
$y$-position, pressure). Hantrakul previously presented an RNN
LightPad controller~\cite{Hantrakul:2018aa}, however that system used
a deterministic RNN without the ability to predict \emph{when} to play
the LightPad. Unlike that system, IMPS can stop swirling and predicts
when and where to start a new touch, a significant advantage allowing
the performance of novel rhythms as well as gestures.

\subsection{X-Touch Mini}
\label{sec:fader-controller}

\begin{figure}
  \centering
  \includegraphics[width=0.3\textwidth]{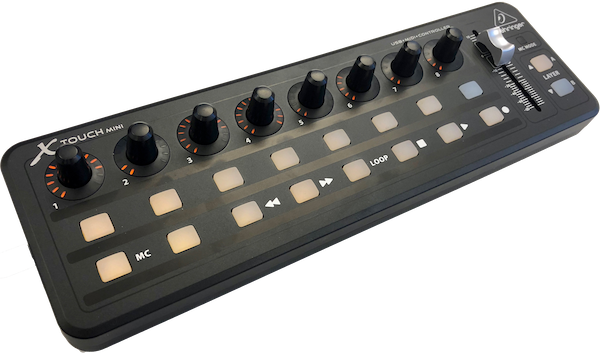}
  \caption{We used the X-Touch Mini to control an additive synthesis
    instrument along with a 9D MDRNN model ($dt$,
    knobs 1--8).}
  \label{fig:x-touch}
\end{figure}

Mixing control surfaces with multitudes of faders and knobs are common
in recording studios as well as in NIME research. Interfaces with illuminated controls, such as
the Behringer X-Touch Mini, can represent previously recorded data as
it is played back (see Figure \ref{fig:x-touch}). We used the X-Touch
Mini to control the parameters of an additive synthesis instrument,
and recorded 10-minutes of performance with this instrument that
comprised 12,000 interactions. This data was used to train a 9D MDRNN
model ($dt$, knobs 1--8). In performance, predictive output can be
sent to a synthesis routine as well as represented visually with the
illuminated controls. ML fader interfaces have previously
been demonstrated without a
neural-network powered controller~\cite{Tahiroglu:2015fj}. Unlike Tahiroglu et al's research,
our system can interpolate between the many training examples
and come up with novel gestures.

\section{Evaluations and Discussion}


\begin{figure}
  \centering
  \includegraphics[width=0.45\textwidth]{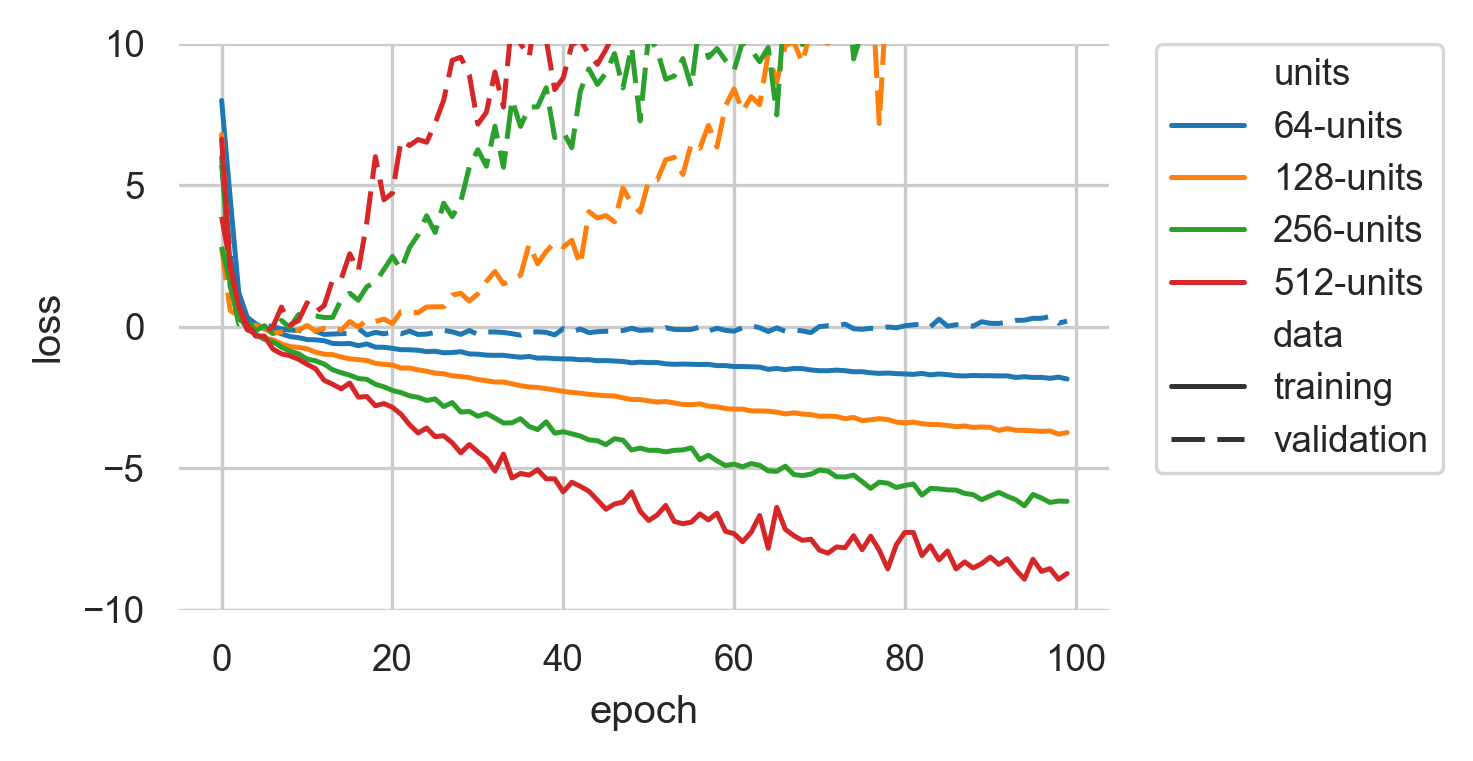}
  \caption{Training and validation loss for a small LightPad
    dataset (12K interactions) for different models
    (lower is better). The 64-unit model had the best performance on
    the validation data.}
  \label{fig:small_dataset_loss}
\end{figure}

\begin{figure}
  \centering
  \includegraphics[width=0.45\textwidth]{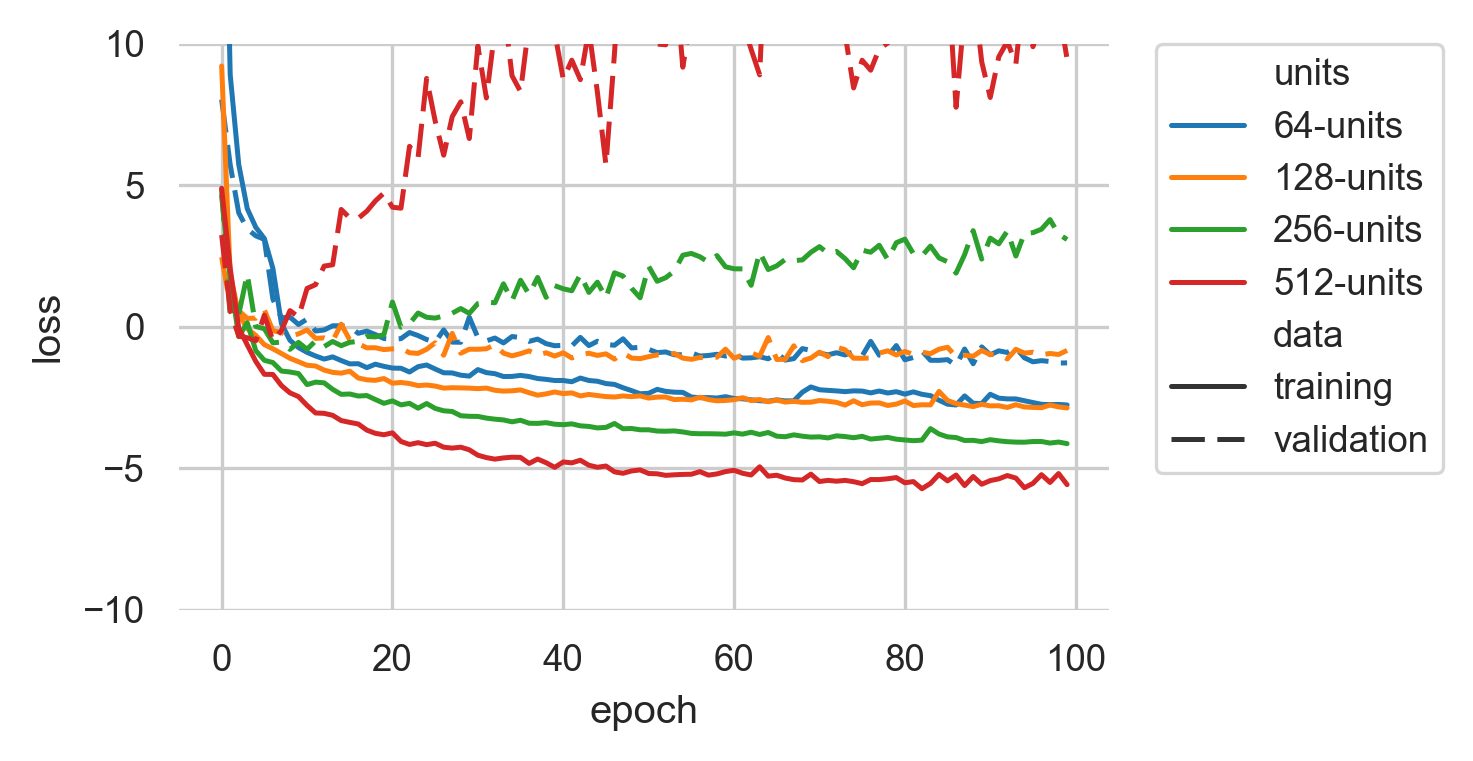}
  \caption{Training and validation loss for a 100K LightPad
    interaction dataset. The 64- and 128-unit models had the best validation
    loss.}
  \label{fig:large_dataset_loss}
\end{figure}

Many aspects of predictive music systems can be evaluated to
determine their place in a NIME context; however, a key question for
NIME designers and performers wishing to apply IMPS would be whether
training is fast enough to keep up with a rapid-prototyping process
and predictions are fast enough to keep up with real-time performance.
In this section we evaluate training performance on small datasets,
prediction speed, and strategies for sampling useful predictions.

We note here that training and inference are very different operations
for the MDRNN. In short, while inference only involves computing the
output from a single timestep, training involves unrolling the MDRNN
to multiple (our default is 50) timesteps to calculate the gradients
of the loss function, this is computed in parallel for each member of
a training batch (our default batch size is 64), thus one training
step requires much more computation than one inference calculation.

\subsection{Training}

Training, as outlined above, is much slower than
inference for DNNs. While training very large DNNs is slow even on
powerful GPUs, the small MDRNNs discussed here can be trained
reasonably quickly on a small dataset, and on a normal computer.
The main hyperparameter to consider here is the size of the MDRNN 
compared to the training dataset; larger MDRNNs have more learning
capacity but are slower to train and may not learn more from a smaller
dataset. By way of example, we consider the LightPad model discussed
in Section~\ref{sec:lightpad-block}. We trained LightPad models on two
datasets: a very small corpus of 12K touch interactions (15 minutes
of performance), and a larger corpus of 100K interactions (2 hours).
The small corpus takes 70 minutes to train for 100 epochs (exposures
to all examples) on a MacBook Pro.


Would it help in this case to use a larger network? We trained 64-,
128-, 256-, and 512-unit networks on each dataset, and used the loss
values to investigate learning performance as shown in Figure
\ref{fig:small_dataset_loss} and \ref{fig:large_dataset_loss}. The
training loss is the average loss over each batch of training
examples, and the validation loss is calculated on a set (10\% in our
case) of examples that are held out to evaluate training after each
epoch. Figure \ref{fig:small_dataset_loss} shows that all but the
64-unit network had poor performance against the validation set on the
12K dataset indicating that the larger networks had overfit to the
training data. For the 100K dataset, Figure
\ref{fig:large_dataset_loss} shows that both the 64- and 128-unit
networks had reasonable validation performance. It is likely that the
smaller networks are most useful for datasets of this size. Early
stopping (an option in IMPS) would have saved much time here by
stopping training after the validation loss failed to improve.

While 70 minutes of training time is more than the mere seconds
required to train Wekinator models~\cite{fiebrink2009metainstrument},
we argue that it does not preclude rapid NIME-prototyping. Similarly
to Wekinator, an MDRNN with our system can be improved by subsequent
retraining. Our predictive interaction system logs all interactions to
facilitate the accumulation of datasets as a NIME is refined. In our
process, models are continually retrained during NIME creation as more
data becomes available through experimentation and performance.


\subsection{Prediction Speed}

We benchmarked prediction speed from our MDRNN on four computer
systems: a MacBook Air (\emph{mba}, Intel i5 1.8GHz), MacBook Pro (\emph{mbp}, Intel
i7 2.6GHz), desktop PC (\emph{gpu}, NVidia GeForce GTX 1080TI), and a
Raspberry Pi 3 B+ (\emph{rpi}, Broadcom BCM2837B0 1.4GHz). The desktop PC
computed MDRNN predictions on the GPU and the other systems used their
CPUs. We examined two aspects of the MDRNN
configuration---the input/output dimension of the network (between 2
and 9), and the number of LSTM units in each RNN layer (64, 128, 256,
512). 100 predictions were performed with each configuration on each
system with the first of each test discarded due to time taken for
setup overhead.

\begin{figure}
  \centering
  \includegraphics[width=0.45\textwidth]{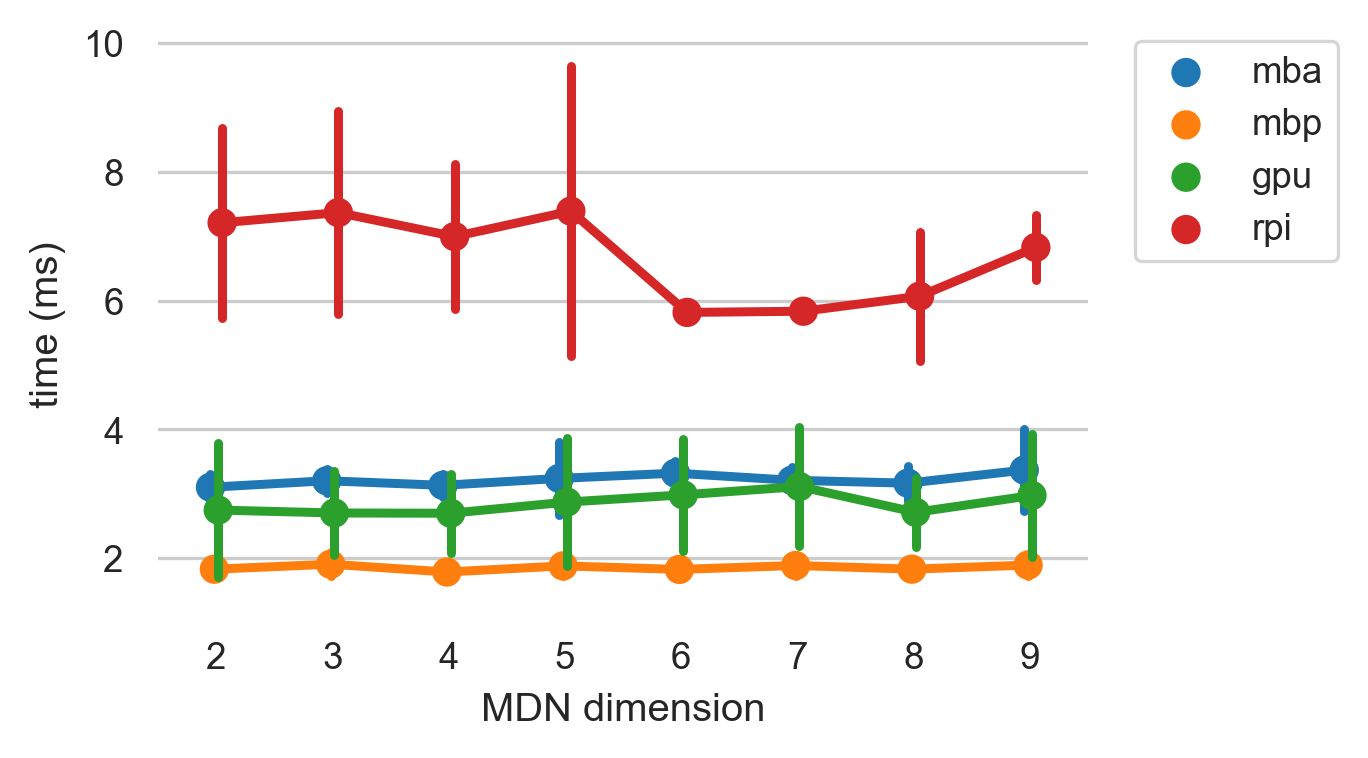}
  \caption{Time (mean and SD) taken per prediction for different MDN
    output dimensions (LSTM units = 64) showing little effect on
    computation speed.}
  \label{fig:prediction-time-dimension}
\end{figure}

The prediction time for a 64-unit per layer MDRNN with different output
dimension sizes is shown in Figure
\ref{fig:prediction-time-dimension}. This shows that changing the
output dimension from 2 up to 9 has little impact on computation speed
for each computer system. The final MDN layer uses few parameters in
comparison to the RNN layers in the network, so its size has little
impact on computation time. A practical consequence is that it is
feasible to experiment with MDRNN prediction of complex NIMEs with
many dimensions of real-valued input.

\begin{figure}
  \centering
  \includegraphics[width=0.45\textwidth]{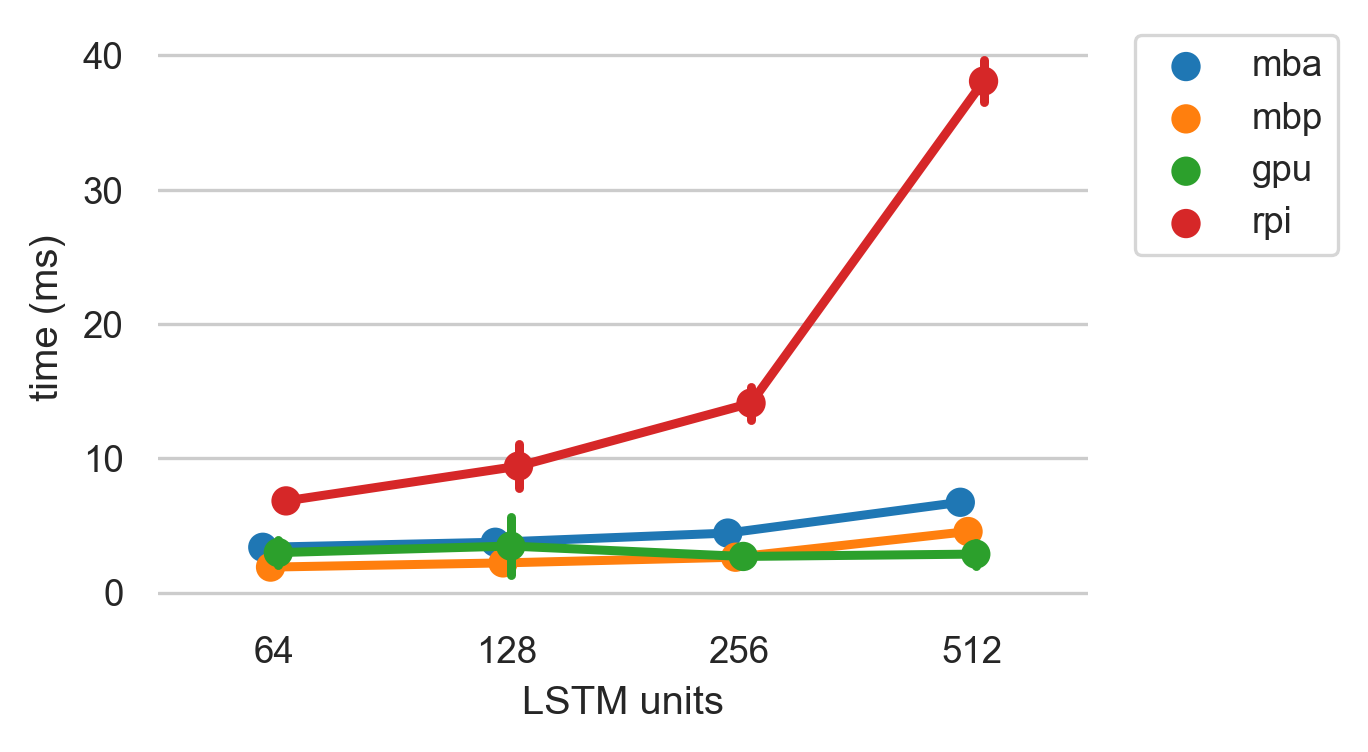}
  \caption{Time (mean and SD) taken per prediction for different RNN
    layer sizes. The time is <10ms for 64-unit networks on all
    systems.}
  \label{fig:prediction-time-units}
\end{figure}

The number of LSTM units in each RNN layer has a much more significant
impact on the speed of prediction (Figure
\ref{fig:prediction-time-units}). The \emph{mbp} achieved the best
speed for the 64 unit network with a mean of 1.85ms per prediction and
a mean of only 4.51ms on the 512 unit network. This means that MDRNN
predictions could fit within typical expectations for NIME latency of
around 10ms~\cite{McPherson:2016aa} on this system. For \emph{gpu},
prediction speed barely changed between 64- (mean=2.85ms) and 512-unit
(mean=2.96ms) networks due to the advantages of GPU-acceleration for DNN
calculations. The \emph{rpi} saw the greatest changes in speed, from a
mean of 6.69ms on the 64-unit network to 37.64ms on the 512-unit
network. While smaller networks are feasible on the \emph{rpi}, larger
networks may be too slow in practice.

\subsection{Sampling}

The output from an MDN is the set of parameters for the GMM: the
weights for each mixture component ($\boldsymbol{\pi}$), and the means
and scales for each multivariate Gaussian ($\boldsymbol{\mu}$ and
$\boldsymbol{\sigma}$). Predictions must be generated from these
parameters by sampling from the categorical distribution formed by
$\boldsymbol{\pi}$, and then sampling from the Gaussian distribution
chosen by that outcome.

The concept of adjusting the temperature of a categorical distribution
will be familiar to those who have used RNNs to learn creative
sequences such as text or symbolic music. The MDN's categorical
distribution can be adjusted in the same way with very low temperature
values favouring the maximum value in the distribution, and high
values producing a more uniform distribution. This adjustment could be
called ``$\pi$-temperature''. The temperature of the Gaussian
distributions can also be adjusted by scaling the variances
$\boldsymbol{\sigma}$. High values result in a wider spread of
predictions, and low values are closer to the selected mean. We call
this ``$\sigma$-temperature''.

\begin{figure}
\centering
\includegraphics[width=0.45\textwidth]{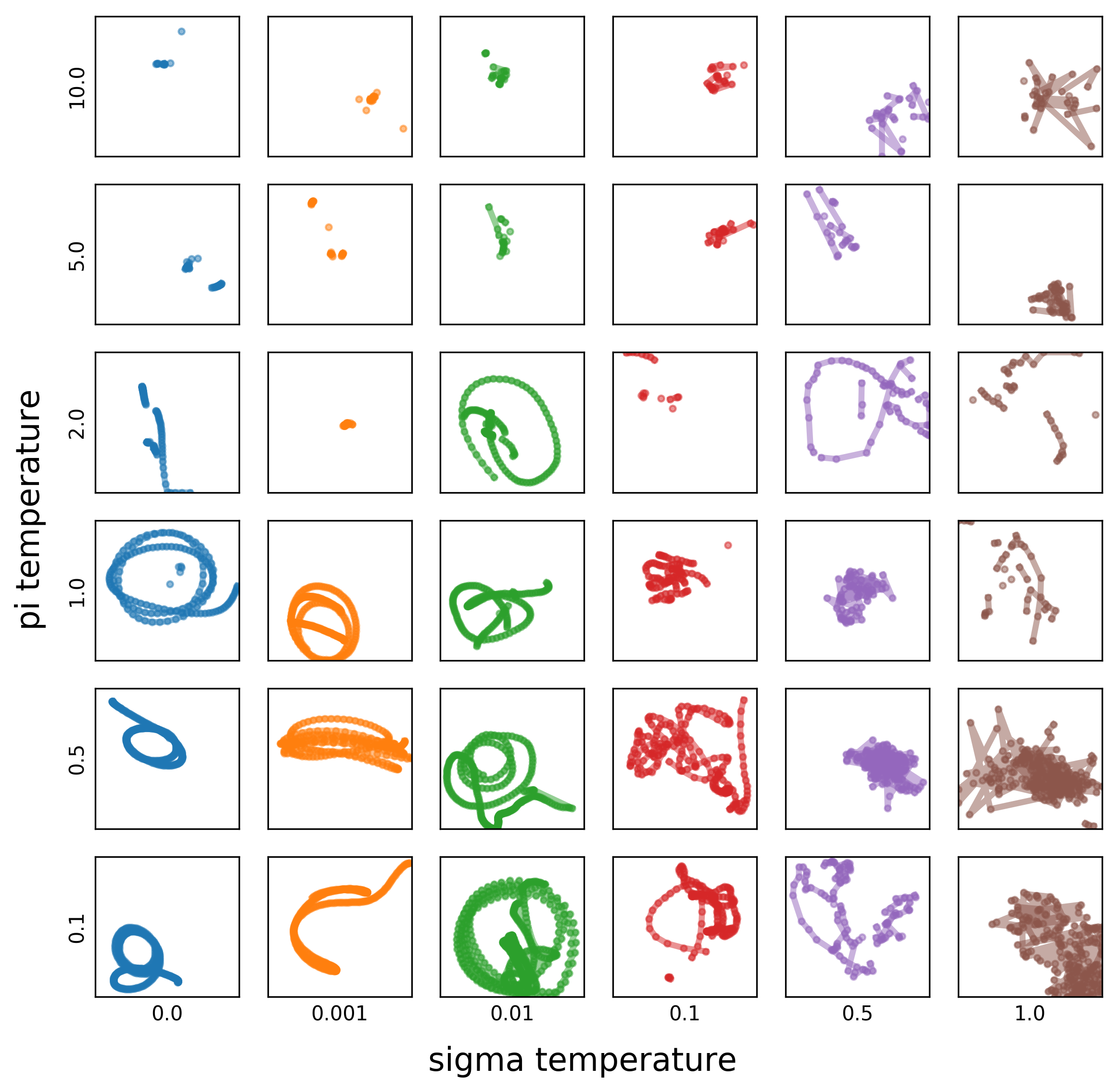}
\label{robojam-temp-sampling}
\caption{Performances from a 3D ($dt, x, y$) MDRNN at different sampling
  temperatures.}
\end{figure}

In IMPS, we have found adjusting the $\sigma$- and $\pi$-temperature
to be very important for making useful predictions. Figure
\ref{robojam-temp-sampling} shows touchscreen performances on a 3D
($dt, x, y$) network generated at different temperatures. The MDRNN
can end up producing high $\boldsymbol{\sigma}$s, resulting in jagged
output, but by reducing the $\sigma$-temperature to close to zero, we
can generate smooth paths. The $\pi$-temperature can control the
appearance of different gestures to some extent. At low values, an
MDRNN will have trouble changing modes, endlessly swirling. At very
high values, it taps without completing any significant swipes.
Exploring how $\pi$ and $\sigma$ sampling temperature can be explored
in our predictive interaction systems is a topic of our future
research.

\section{Conclusion and Future Work}

We have introduced IMPS, an Interactive Musical Prediction System,
that allows an MDRNN model to be applied in any NIME with continuous
control signals. This system has been demonstrated through predictive
interactions with three interfaces: the custom built EMPI, 
commercial LightPad, and X-Touch Mini. Our evaluation has shown that
IMPS is viable for use on normal laptops and even single-board
computers, and provided insight into how the MDRNN behaves under
different training and sampling parameters. It is promising that
small, easily trained models seem sufficient to learn smaller
datasets; however, future studies could seek to understand what aspects of a
control style are actually learned and how IMPS might work within a
design process. 

We think that predictive interaction could be applied widely in
computer music software such as DAWs, synthesis environments and
physical hardware controllers. The success of tools such as
Wekinator~\cite{Fiebrink:2017aa}, and interest in Google's Magenta
project suggest that artists see the value of applying ML in their
work. The flexibility of IMPS and our MDRNN design could be ideal for
providing predictive interaction possibilities to these users.

\subsubsection*{Acknowledgements}
\anonymize{This work is supported by The Research Council of Norway as part of
the Engineering Predictability with Embodied Cognition (EPEC) project
\#240862 and the Centres of Excellence scheme, project \#262762.}

\bibliographystyle{abbrvurl}
\bibliography{references.bib}

\end{document}